\documentclass[journal=jacsat,manuscript=article]{achemso}

\usepackage{chemformula} 
\usepackage[T1]{fontenc} 
\usepackage{graphicx}
\usepackage{comment}
\usepackage{hyperref}
\usepackage{algorithm}
\usepackage{algorithmic}
\usepackage{wrapfig}
\usepackage{amsmath}
\usepackage{mathrsfs}
\usepackage{amsfonts}

\author{Ayana Ghosh}
\email{ghosha@ornl.gov}
\affiliation[1]
{Computational Sciences and Engineering Division, Oak Ridge National Laboratory, Oak Ridge, TN  37831, USA}

\author{Palanichamy Gayathri}
\affiliation[SRMISTNANO]
{Department of Physics and Nanotechnology, Faculty of Engineering and Technology, SRM Institute of Science and Technology, Kattankulathur - 603 203, Tamil Nadu, India}

\author{Sathiyamoorthy Buvaneswaran}
\affiliation[SRMISTNANO]
{Department of Physics and Nanotechnology, Faculty of Engineering and Technology, SRM Institute of Science and Technology, Kattankulathur - 603 203, Tamil Nadu, India}

\author{Saurabh Ghosh}
\email{saurabhghosh2802@gmail.com}
\affiliation{Department of Physics and Nanotechnology, Faculty of Engineering and Technology, SRM Institute of Science and Technology, Kattankulathur - 603 203, Tamil Nadu, India}
\alsoaffiliation{Center for Advanced Computational and Theoretical Sciences, SRM Institute of Science and Technology, Kattankulathur - 603 203, Tamil Nadu, India}

\title[An \textsf{achemso} demo]
{Intervention strategies for polarization switching in hybrid improper ferroelectrics}

\begin{document}

\begin{abstract}
{{\color{black}
The potential of hybrid improper ferroelectrics (HIFs) in electronic and spintronic devices hinges on their ability to switch polarization.
Although the coupling between octahedral rotation and tilt is well established, the factors that govern switching barriers remain elusive.
In this study, we explore this area to demonstrate the critical role of causal reasoning in uncovering the mechanisms to control the ferroelectric switching barrier in HIFs. 
By combining causal discovery, causal interventions, and first-principles simulations, we identify tolerance factor, A-site cation radii mismatch, epitaxial strain, and octahedral rotation/tilt as key parameters and quantify how their interplay directly influences switching barrier. 
Three key insights emerge from our work: (a) the analysis identifies the structural descriptors controlling polarization reversal across a broad family of A-site-layered double perovskites and superlattices, (b) it uncovers non-trivial, material-specific rotation-tilt mechanisms, including a counterintuitive cooperative pathway where both rotation and tilt change while lowering the barrier, an effect mostly inaccessible to conventional Landau or first-principles-based approaches and (c) it maps these material-specific mechanisms to experimentally realizable parameters, showing that epitaxial strain from orthorhombic substrates (e.g., NdScO$_3$, NdGaO$_3$) selectively tunes octahedral distortions to achieve barrier reduction across varied compositions.
These results establish actionable, materials-by-design principles linking composition, structure, and strain to polarization switching, while highlighting the potential of causal reasoning to guide intelligent, mechanism-driven strategies for engineering complex functional oxides.}}
\end{abstract}

\section{Introduction}
Ferroelectric materials continue to be an active area of research even after a hundred years of its first discovery in Rochelle salt \cite{Valasek1921}.
Spontaneous symmetry breaking, hysteresis and domain structure are the primary working physical principles behind the exhibition of ferroelectricity \cite{PhyFERevModPhys}. 
This functionality can manifest in three main types: proper, improper, and hybrid improper ferroelectricity.
Proper Ferroelectricity (PF) occurs when the spontaneous polarization is a direct result of a structural phase transition \cite{acosta2017batio3}.
Polarization is intrinsic to the crystal structure of a material and originates from the shift in symmetry from non polar to polar phase.
%
\begin{figure*}
\centering
\includegraphics[width=\textwidth]{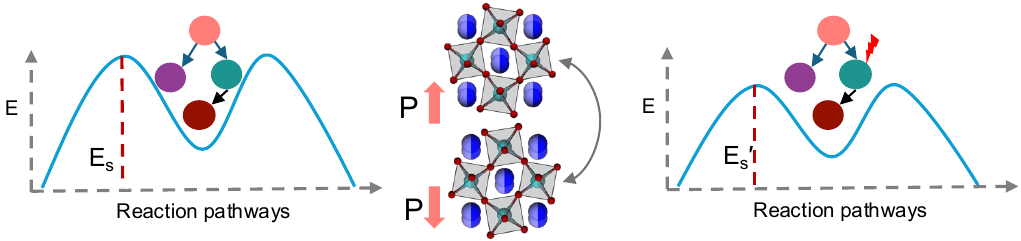}
\caption{Outline of how causal interventions can elucidate the fundamental atomistic mechanisms governing ferroelectric switching in hybrid improper ferroelectric oxides.
}
\label{fig:overview}
\end{figure*}
In improper ferroelectricity (IF), the polarization arises indirectly as a secondary effect of a primary phase transition that is driven by some other order parameter, such as a magnetic or structural distortion \cite{bousquet2008improper, SNAFM}.
Spontaneous polarization is not the primary driving force but rather a consequence of another phase transition.
{\color{black}
The third kind is hybrid improper ferroelectricity (HIF) where polarization arises from the trilinear coupling of two or more non-polar structural distortions. 
Polarization in this scenario is governed by the cooperative interplay of multiple distortions, neither relying on polarization as the primary order parameter, as in proper ferroelectrics, nor arising from a single primary distortion, as in improper ferroelectrics.
}
In HIFs, the polarization results from combining two or more structural distortions that are not individually polar but collectively induce spontaneous polarization \cite{benedek, HIFexperimental}.
These distinctions are crucial for understanding the diverse mechanisms leading to ferroelectric behavior in various materials.
Traditionally, studies have explored numerous systems such as perovskite oxides \cite{MZHIFSW, mulder2013turning}, halides \cite{zhang2020halide}, metal-organic frameworks \cite{stroppa2013hybrid, ghosh2015strain} organic crystals \cite{horiuchi2008organic} through simulations and characterization experiments, with recent emphasis on interpreting physical insights from data-driven machine learning (ML)/ deep learning (DL) approaches \cite{GPswCM2,ghosh2022insights}.
\par
Our current investigation aligns closely with this domain, focusing on ferroelectric switching in HIF oxides which is one of the key attributes for advancing our fundamental understanding of the spintronics-based device suitability \cite{eerenstein2006multiferroic, spaldin2005renaissance, PhyFERevModPhys, waser2009redox, garcia2014ferroelectric, scott2007applications, GPswCM1} of these materials.
Here, the switching \cite{PhyFERevModPhys, BTOPSW, danielsHIF, GPswCM1, GPswCM2, MZHIFSW, TBHIFSW, mulder2013turning} refers to the reversal of electric polarization in ferroelectric materials when subjected to an external electric field, overcoming a potential energy barrier, known as the switching barrier (E$_\text{s}$).
The rearrangement of the coupled distortions under an applied electric field \cite{TBHIFSW} in HIFs lead to complex switching dynamics, including multi-step switching in addition to the existence of multiple polarization states \cite{GPswCM1, GPswCM2}.
\par
{\color{black}
Similar multi-step domain switching mechanisms have been reported in layered Ruddlesden-Popper HIFs \cite{nowadnick2016domains,oh2015experimental}, highlighting the broader relevance of structural coupling in governing switching behavior.
For example, a previous study by Nowadnick et al. \cite{nowadnick2016domains} on the Ruddlesden–Popper phase system Ca$_3$Ti$_2$O$_7$, a prototypical HIF material, combined group-theoretical analysis with first-principles calculations to enumerate low-energy ferroelectric switching pathways and domain structures. 
The study showed that polarization reversal proceeds via multiple routes, involving coupled changes in octahedral rotation and tilt modes as well as antipolar distortions.
\par
In general, the switching process is highly complex, as it involves a web of interdependent parameters that do not always conform to conventional mechanisms via the primary order parameters \cite{benedek,mulder2013turning}.
Although first-principles calculations combined with phenomenological modeling provide valuable insights, they are often limited in identifying mechanisms that generalize across compositions or in suggesting actionable strategies to tune the switching barrier (E$_\text{s}$). 
Identifying parameters that directly correspond to tunable factors, those that could serve as guiding principles for controlling E$\text{s}$ remains an open problem (Figure~\ref{fig:overview}).
\par
These challenges naturally underscore the need for data-driven, artificial intelligence/machine learning (AI/ML)-guided strategies \cite{MLjordan2015machine, MLlinardatos2020explainable,DLlecun2015deep, MLDLjaniesch2021machine} that can systematically uncover mechanism-informed, experimentally relevant design principles.
However, reliance on \textit{data correlations, pattern recognition} underneath existing data, lack of explainability, interpretability \cite{AIwang2023scientific} and more importantly reasoning abilities \cite{CIprosperi2020causal} of the AI/ML models pose hindrance to understand the underlying physics or chemistry governing the material properties.
In other words, while most implementations of ML/DL models can effectively identify patterns and correlations in data (association), they often fall short in providing insights into how material parameters influence properties (intervention) or predicting what would occur under alternative, unobserved conditions (counterfactual). 
\textit{Consequently, a rigorous solution to the switching-barrier problem demands the incorporation of a modern causal-reasoning methodology.}
Without it, identifying the key mechanisms needed to inform E$_\text{s}$ control strategies becomes exceedingly difficult, if not altogether impossible.
\par
In this paper, we address this knowledge gap across a broad chemical space of A-site–layered double perovskites (DPOs) and superlattices (SLs).
We introduce a causal reasoning approach to guide the decision-making process followed by performing physics simulations to validate the outcomes. 
It enables us to (a) construct causal discovery models to identify the underlying cause-effect relations between electronic structure, geometry-based factors and E$_\text{s}$ for a variety of HIF DPOs (AA$^\prime$BB$^\prime$O$_6$) and SLs (ABO$_3$/A$^\prime$BO$_3$), (b) down-select features based on physics knowledge and fine-tune the model supported by refutability tests, (c) perform interventions while adhering to physics-constraints to obtain desired E$_\text{s}$ and (d) elucidate causal mechanisms in conjunction with their connections to outcomes from physics simulations to yield experimentally-accessible design principles to tune E$_\text{s}$.
The key steps of the approach is illustrated in Figure~\ref{fig:causal models}.
%
\begin{figure*}
\centering
\includegraphics[width=\textwidth]{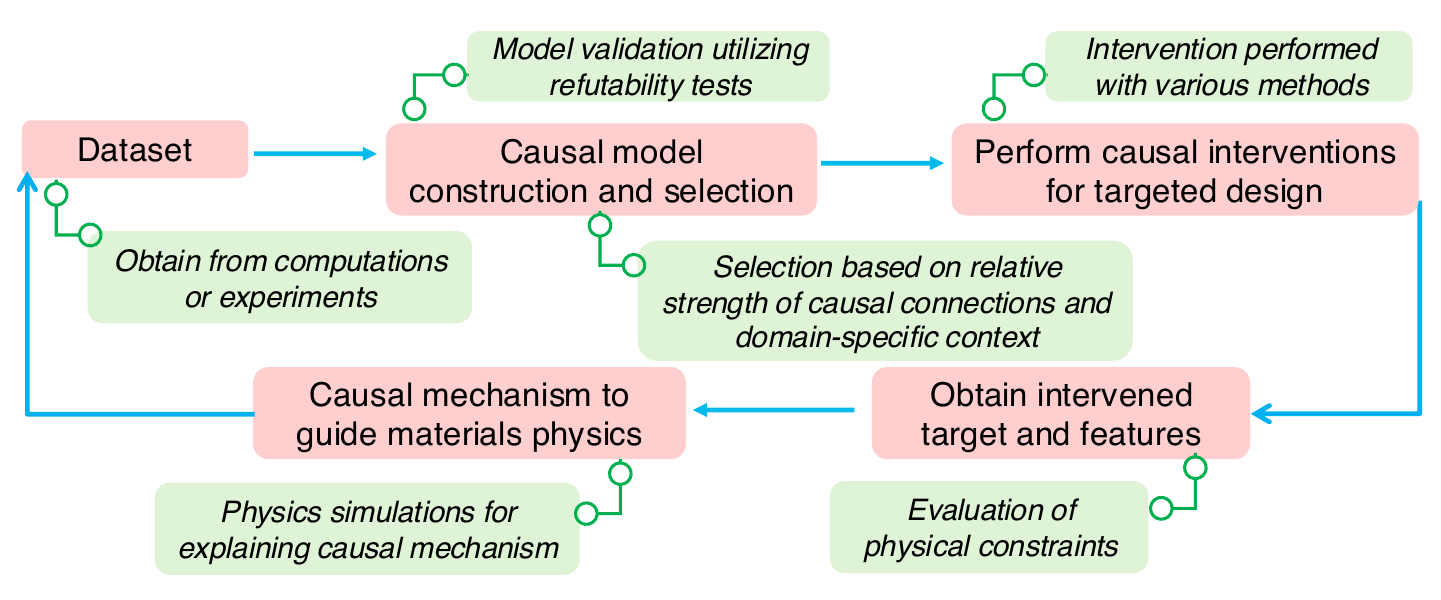}
\caption{Key steps in integrating causal modeling with physics simulations to derive mechanisms for reducing switching barriers. The approach employs algorithms for causal modeling to address design tasks effectively, followed by causal interventions that incorporate physics-based constraints to gain insights into the underlying processes guiding the phenomena.
}
\label{fig:causal models}
\end{figure*}
%
\par
Our results reveal several central insights into the mechanisms governing the barrier.
Octahedral rotation, tilt, tolerance factor, and A-site size mismatch each have quantifiable, material-specific causal effects on E$_\text{s}$.
Targeted interventions show how controlled adjustments to these descriptors lower the barrier within physically allowable limits. 
Three distinct rotation–tilt pathways emerge: two consistent with Landau-type \cite{mulder2013turning, TBHIFSW, GPswCM1, GPswCM2, MZHIFSW} expectations, and a third, counterintuitive mechanism where simultaneous increases in rotation and tilt cooperatively reduce E$_\text{s}$.
The causal mechanisms point to an underlying variable capable of driving the required structural changes, which can be realized experimentally through epitaxial strain.
Mapping the intervention-predicted structural changes onto achievable tensile or compressive strain states from orthorhombic substrates, and validating the outcomes with first-principles simulations, confirms that cooperative distortion tuning can flatten the polarization-reversal landscape even when both nonpolar distortions increase. 
These findings yield experimentally accessible, mechanism-grounded, actionable design rules that link composition, structural descriptors, and substrate choice to the tuning of barrier in HIFs.}
\par
{\color{black}
We note that the switching pathways analyzed in our study are restricted to coherent polarization reversal, in which the order parameters evolve uniformly without the formation or motion of domain walls. 
Widely adopted in first-principles studies, this approach enables consistent comparison of intrinsic switching barriers across varied compositions. However, it does not capture the full complexity of experimental switching, where extrinsic processes such as domain wall nucleation, motion, together with defect interactions often dominate the macroscopic coercive field. 
Our results should therefore be interpreted as intrinsic lower bounds to the energy cost of polarization reversal, rather than as direct predictions of experimentally measurable switching fields. 
Similar distinctions between coherent switching and more realistic experimental pathways have been discussed in prior studies \cite{beckman2009ideal,nowadnick2016domains,munro2018discovering,munro2019implementation}. 
Thus, the primary contribution of this study lies in the identification of material-specific causal mechanisms based on the interplay of octahedral rotation and tilt distortions along with their tunability by strain that control intrinsic switching barriers. These lead to identification of experimentally accessible design rules for polarization switching for HIFs.}
\section{Methods}
\subsection{Details on DFT computations}
We have considered 159 A-site layered DPOs as available from the \href{https://zenodo.org/record/6570994}{publicly available repository} from our previous studies \cite{GPswCM1, GPswCM2}. 
A total of randomly selected 133 non-magnetic SLs (in $Pmc2_1$ symmetry group) is considered in this study, curated from 465 ABO$_3$/A$^\prime$BO$_3$ SLs falling in different zones based on criteria of ionic radii mismatch and $\tau$. 
\par
The first-principles calculations are performed by employing density functional theory (DFT)\cite{DFT} with projector augmented wave (PAW) potentials \cite{joubertPAW} utilizing the generalized gradient approximation \cite{PerdewGGA} with effective Hubbard correction ($U_{\text{eff}}$)\cite{LDAU} within the Vienna \textit{ab initio} simulation package (VASP) \cite{hafner1997vienna}.  
The exchange-correlation component is approximated using the PBEsol functional\cite{PBESol}. 
The computations are based on a $\sqrt{2}a \times \sqrt{2}a \times 2c$ pseudo cubic cell containing a total of 20 atoms. 
Brillouin zone integrations are performed using a $\Gamma$-centered 6$\times$6$\times$4 $k$-point mesh in accordance with crystal symmetry. 
We set a cutoff energy of 520 eV for all calculations, incorporating spin polarization. 
Geometry relaxations continue until changes in total energy between relaxation steps are within 1$\times$10$^{-6}$ eV, and atomic forces on each atom are below 0.01 eV/Å. 
G-type antiferromagnetic (AFM) ordering with collinear spins at B/B$^\prime$ sites is assumed for all computations. 
To account for on-site Coulomb interaction for 3$d$ states of transition metals, we employ an effective Hubbard parameter (U$_{\text{eff}}$ = U - J$_{H}$), with specific values for V (2.0 eV), Cr (3.0 eV), Mn (4.0 eV), Fe (4.5 eV), Co (5.0 eV), Ni (6.0 eV), as well as 4$d$ states of transition metals including Mo (4.5 eV), Zr (0 eV), Ru (5.0 eV), and 5$d$ states such as W (0.0 eV), Re (1.0 eV), Os (1.0 eV), respectively. 
The selection of U$_{\text{eff}}$ is guided by ensuring the representation of correct oxidation states, as evidenced by the accurate estimation of local magnetic moments at the BB$^\prime$ sites and electronic structure in the DPOs and SLs and also extensively 
justified in your previous work on DPOs \cite{ghosh2022insights, GPswCM1, GPswCM2}.
The E$_\text{s}$ for all compounds in the dataset are estimated using Nudge Elastic Band (NEB) method as implemented in VASP \cite{jonsson1998nudged}.
The end-point structures for DPOs considered are the fully distorted structures belonging to $P2_1$ and $P4/n$ space group symmetry.
The structure with $P4/n$ symmetry corresponds to the Q$_{\text{R-}}$ distorted phase.
The low-symmetry structure for SLs belongs to the $Pmc2_1$ space group, while the phase with the Q$_{\text{R-}}$ distortion is associated with the $P4/nbm$ space group.
The intermediate phase is generated by interpolating between the initial and final stages, identified as the configurations with localized minima acquired through relaxation.
Strain is introduced in orthorhombic DPOs by constraining the in-plane lattice constants to the substrate lattice parameters, while allowing relaxation along the $c$-axis to achieve volume optimization.
%
\begin{figure*}
\centering
\includegraphics[width=\textwidth]{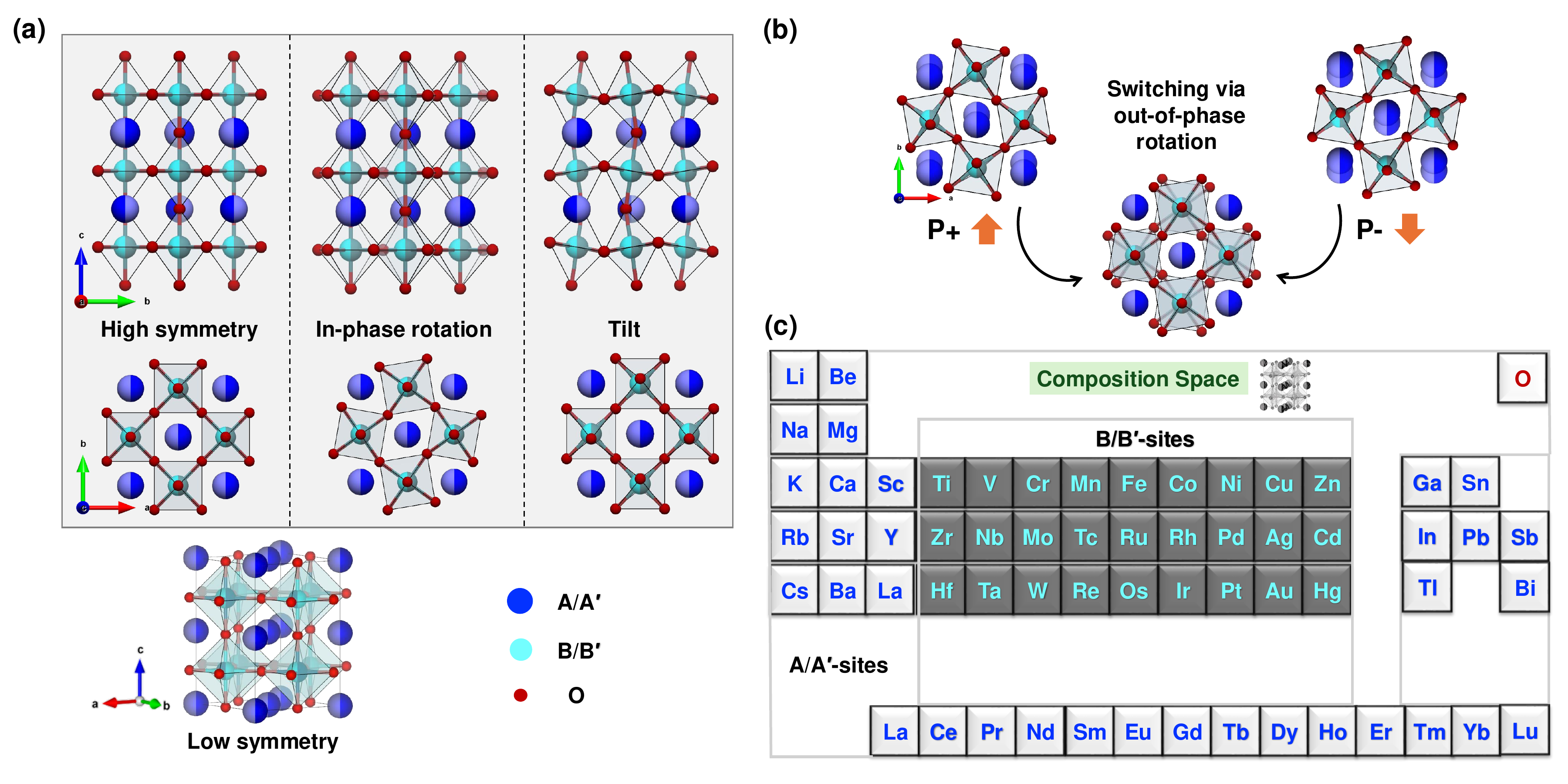}
\caption{Schematic representation of the (a) high symmetry phase, key structural modes such as in-phase rotation, tilt in superlattices (upper panel), double perovskite oxides (middle panel) and low symmetry phase (bottom panel). (b) The structural distortions corresponding to polarization switching via out-of-phase rotation of the BO$_6$ octahedra. (c) Illustration of the compositional space showcasing the chemical diversity of cations, as considered in the study.}
\label{fig:modes}
\end{figure*}
%
{\color{black}
\par
Our current dataset is mostly restricted to DPOs and SLs in which A-site layered ordering is either experimentally observed or theoretically predicted to be favorable. 
This selection is based on a central design principle for stabilizing HIF in perovskite oxides, i.e., the formation of A/A$^\prime$ cation ordering. 
Such ordering breaks inversion symmetry and facilitates the coupling between octahedral rotations and tilts. 
For chemically distinct A-site cations, layered ordering also enhances the trilinear coupling that drives polarization.
\par
Previous studies \cite{anderson1993b,king2010cation,vasala20151} have demonstrated that the stability of this ordering is influenced by factors such as ionic size mismatch, charge differences, and tolerance factors.
More recent foundational work by Ghosh et al. \cite{ghosh2022insights, GPswCM2}, combining first-principles calculations with causal modeling, has shown how specific structural distortions directly influence the formation of layered ordering. These insights have led to actionable structural guidelines to design novel layered DPOs and SLs.
Building on this body of work, our study focuses on polarization switching in these material systems, where the presence of favorable cation ordering provides a robust structural guideline for realizing and tuning HIF. By focusing on these chemically favorable systems, we ensure that the causal insights derived from our study can be directly relevant to experimentally realizable materials.
}
\subsection{Details on causal reasoning workflow}
Structural causal or equation models  \cite{causaldavidson1967causal, bunge2017causality, shimizu2011directlingam, hyvarinen2013pairwise} informed with prior physics knowledge offer the potential to uncover hidden factors that may not be readily apparent from the data alone, offering insights and reliable pathways for targeted design.
By anchoring decision-making in the fundamental cause-effect relations rather than solely on correlations, causal models offer a more robust pathway for advancing scientific understanding and facilitating informed materials design.
These models face challenges in adoption within physical sciences, except within a few studies\cite{ghosh2022insights, sghosh2024insights, li2024diverse, causalaghosh2024towards} due to a number of factors including system complexity, modeling uncertainty, integration with physical laws, computational demands. 
Despite these hurdles, recognition of their potential benefits in fields such as systems biology
\cite{bioamundson1994function, bioross2021causal}and environmental\cite{adams2003establishing}science, health care \cite{CIprosperi2020causal} have only continued to grow in recent years. \\
\textbf{Initial model construction:}
The primary causal discovery algorithm we utilize is the linear non-Gaussian acyclic model (LiNGAM).
Given a dataset $\textbf{X}$ = $[X_1, X_2, ..., X_n]$, the first goal in the workflow is to learn the causal structure by estimating an adjacency matrix $\textbf{A}$ representing causal dependencies between the variables.
Note that $\textbf{X}$ represents the input matrix of features in the dataset for different samples.
The prior knowledge constrains the model by specifying known causal relationships (e.g., which variables are exogenous or sinks).
The algorithm first estimates the causal order of the variables, ensuring that each variable is conditionally independent of its predecessors. 
After determining the causal order, the coefficients $a_{ij}$ of the causal relationships are estimated by minimizing the non-Gaussianity of the residuals:
\begin{equation}
X_i = \sum_{\substack{j \neq i}} a_{ij} X_j + \epsilon_i
\end{equation}
The residuals $\epsilon_i$ are minimized based on a non-Gaussianity measure to determine the causal coefficients.
The LiNGAM (Direct) model is then fitted to the data $\textbf{X}$.
The evaluated adjacency matrix $\textbf{A}$ represents the learned causal structure.\\
\textbf{Model selection and visualization:}
Once the model is learned, the adjacency matrix $\textbf{A}$ is used to visualize the causal relationships. 
An edge from $X_j$ to $X_i$ is drawn if the absolute value of the corresponding coefficient $a_{ij}$ exceeds a threshold $\epsilon$ with the edge labeled by $a_{ij}$.
This is represented by the directed acyclic graph (DAG) where the nodes are variables, and edges are the causal dependencies between them.
To simplify the model, features with relatively weak causal relationships are removed by setting coefficients below a user-specified threshold $\tau$ to zero.
The threshold is determined by the selected percentile of the absolute values in the adjacency matrix.
The updated adjacency matrix $\textbf{A}_\text{filtered}$ is computed as $\textbf{A}_\text{filtered}$ = $\textbf{A}[|a_{ij}|] \geq \tau]$, where $\tau$ is the user-defined threshold for the coefficient's magnitude.
In addition, depending on the design tasks, the user is given the choice to manually select features to carry out the down-selection.
\\
\textbf{Refutability tests:} The following falsifiability tests are designed to evaluate the robustness of causal models by examining their consistency and stability under various perturbations and shifts.\\
\underline{Test I: Causal model structure consistency}\\
In this test, the goal is to evaluate whether the causal model structure remains consistent across iterations when slight perturbations are made to the feature values based on their statistical properties (mean and standard deviation).
For each feature $f_i$ in the dataset $\textbf{X}$, we apply the following perturbation based on its statistical properties (mean $\mu_i$, standard deviation $\sigma_i$):
\begin{equation}
{X_i}^\text{perturbed} = X_i + \epsilon_i
\end{equation}
where $\epsilon_i$ $\sim$ $\mathscr{U}(-\Delta_i, \Delta_i)$ with $\Delta_i$ = abs($X_i$ - $\mu_i$).
Here, $\mathscr{U}$ denotes a uniform distribution and $\Delta_i$ represents the range of perturbation for feature $X_i$.
Using the perturbed dataset, the causal model is fit to obtain a new adjacency matrix.
Within each iteration, the adjacency matrices of the original and perturbed datasets are checked to evaluate if they are consistent within a tolerance of $\epsilon_\text{adj}$.
If the normalized difference between the two is below the tolerance, then the model passes the test.\\
\underline{Test II: Causal model order consistency}\\
This test checks whether the causal ordering of variables is consistent across iterations when slight perturbations are applied to the feature values.
Similar to the previous test, for every iteration the adjacency matrices corresponding to the original versus perturbed datasets are obtained within each iteration. 
Here, the matrix sizes/orders care compared and if they are the same, then the model passes the test.\\
\underline{Test III: Edge stability}\\
This test examines whether the causal edges (the connections between variables) remain stable when small random noise is added to the dataset.
Both Gaussian and non-Gaussian noises can be provided.
Small noises are added to the dataset :
\begin{align}
{X_i}^\text{perturbed} &= X_i + \mathscr{N}(0,\sigma^2), \notag \\
\text{Or,} \notag \\
{X_i}^\text{perturbed} &= X_i + \mathscr{L}(0,b)
\end{align}
where $ \mathscr{N}(0,\sigma^2)$ represents Gaussian noise and $\mathscr{L}(0,b)$ denotes Laplace noise.
Within each iteration, the adjacency matrices of the original and perturbed datasets are checked to evaluate if they are consistence within a tolerance of $\epsilon_\text{edge}$.
If they are found to be similar within a given tolerance, then the edges are stable, otherwise unstable.\\
\underline{Test IV: Distributional shift}\\
This test evaluates whether the causal relationships are consistent when a small distributional shift is applied to one of the features.
A small shift to a feature of user's choice is applied by adding noise to the specified feature.
Let's assume in the original dataset, $\textbf{X}_i$ is the i-th feature.
\begin{align}
{X_\text{shifted}}^\text{new} &= X_\text{shifted} + \mathscr{N}(0,\sigma^2), \notag \\
\text{Or,} \notag \\
{X_\text{shifted}}^\text{new} &= X_\text{shifted} + \mathscr{L}(0,b)
\end{align}
The corresponding adjacency matrices are evaluated, compared to check if they are consistent within a tolerance of $\epsilon_\text{shift}$.
If the normalized difference between the two matrices within each iteration is below $\epsilon_\text{shift}$, then the causal relationships are considered to be stable.\\
Overall, Test I compares the adjacency matrices of the original and perturbed datasets to check if the causal model structure remains consistent across multiple iterations, using small feature modifications.
Test II assesses the consistency of the causal ordering by comparing the adjacency matrices of the original and perturbed datasets, focusing on whether the causal relationships' order remains unchanged.
Test III evaluates whether the edges in the causal model remain stable under small perturbations, with the option of applying either Gaussian or Non-Gaussian noise to the dataset.
Test IV applies a distributional shift to a user-specified feature and checks whether the causal model remains stable by comparing the adjacency matrices of the original and shifted datasets across multiple iterations.\\
\textbf{Interventions:}
Causal intervention methods vary in how they treat the relationships between features.
The choice of intervention mode can significantly influence the results of the analysis.
Keeping these in mind, our intervention setup provides flexibility in how interventions are performed, allowing the user to choose the mode based on whether they are interested in simultaneous, or progressively adjusted interventions. \\
Let's assume the dataset for $n$ samples and $p$ features can be represented by:
\begin{equation}
\mathbf{X} = \left[
\begin{matrix}
x_1^T \\
x_2^T \\
\vdots \\
x_n^T
\end{matrix}
\right]
\end{equation}
where $x_i \in \mathbb{R}^p$ is a vector representing the feature values for the $i$-th sample.
The target is represented as $x_i \in \mathbb{R}^n$.
Given a feature $x_j$, the intervention causal effect can be defined as:
\begin{align}
CE^{(x_j)} = E\left[ \hat{y}(x_j \leftarrow \text{intervention}) \mid X_{-j}, y \right]
\end{align}
Here $\mid X_{-j}$ represents all the features except for $x_j$ and y is the target variable. 
The formulation reflects the expected outcome $\hat{y}$ after performing the intervention on x$_j$.
Given a set of features by the choice of the user, the causal intervention is conducted either in the \textit{joint} or \textit{sequential} manner.
The intervention is bounded within a range defined by the original values of the feature, and it tries to optimize for a target value of the outcome.
The bound takes the domain expertise into account such that the intervention is performed complying with physics-based constraints.
For this specific problem, we have considered a change of up to 35\% for each feature.
The adaptive change within the \textit{sequential} mode is controlled by a weighted threshold coefficient ($\alpha_j$) that is proportional to the original feature value ensuring the intervention is accumulated sequentially.
The intervention bounds then become:
\begin{align}
\text{min\_value}_j^{(k)} = \max \left( x_j^{(k)} - \alpha_j \cdot |x_j^{(k)}|, \min(X_j) \right)
\end{align}
\begin{align}
\text{max\_value}_j^{(k)} = \min \left( x_j^{(k)} + \alpha_j \cdot |x_j^{(k)}|, \max(X_j) \right)
\end{align}
In the \textit{joint} mode, interventions on all selected features are performed simultaneously. 
The causal effects are estimated and adjusted for all the features in the intervention list at once, based on a target value. 
The combined influence of the features on the outcome is considered collectively.
In the \textit{sequential} mode, interventions are applied step-by-step, one feature at a time. 
The changes made in earlier steps influence subsequent interventions. 
This allows for a more structured and gradual refinement of the model. After modifying one feature, the following interventions are adjusted according to the updated context of the dataset.
If no target value is provided, it is sampled from the distribution of the target variable $y$ which can be represented as:
\begin{align}
\hat{y}^k \sim \mathscr{U}(\min(y), \max(y))
\end{align}
where $\mathscr{U}(.)$ represents the uniform distribution.
\par
To account for the variability in causal effects across different runs, we average the results of multiple intervention runs. 
This ensemble approach helps to reduce noise and provides a more robust estimate of the intervention’s impact. 
The results from each run are averaged to yield a final set of interventions, as shown in the following representation :
\begin{equation}
\mathbf{X_\text{intervened}} = \left[
\begin{matrix}
x_1^\text{intervened} \\
x_2^\text{intervened} \\
\vdots \\
x_n^\text{intervened}
\end{matrix}
\right]
\end{equation}
In summary, by adopting either the joint or sequential intervention modes, users can gain flexibility in how they perform causal analysis. 
The ensemble approach (average of multiple runs) further ensures that the results are robust and less sensitive to outliers or randomness in individual runs. \\
\textbf{Mapping to physics simulations:}
In the last part of the workflow, we focus on validating the causal mechanisms identified in the earlier steps (e.g., causal relationships between features and their influence on the target).
We perform checks to ensure the changes in the intervened data lead to physically reasonable changes in the outcome variables.
Causal reasoning is applied to identify and intervene on the key features influencing the target. 
First-principles computations (such as DFT and other physics simulations) are then performed to evaluate the suggested changes in these features resulting from the causal intervention. 
In this process, the causal mechanisms are explicitly modeled, and the computational results help establish links between the predicted outcomes and experimentally achievable pathways. 
This creates a causal reasoning workflow for targeted design, while also offering insights into novel physical mechanisms that can be used to validate and refine the causal reasoning in real-world applications.

\section{Results \& Discussion}
\subsection{HIF ferroelectric mechanism and polarization switching}
In our study, we focus on a dataset of HIF oxides in the form of DPOs and SLs, which were constructed using first-principles computations.
It comprises of a diverse range of combinations of alkaline-earth, rare-earth, lanthanide, and transition metal ions, covering a broad spectrum of charge states, coordination numbers, ionic radii mismatch, tolerance factors ($\tau$).
The ferroelectricity in HIFs originates from the trilinear coupling between multiple order parameters.
The coupling connects two unstable zone-boundary distortions, labeled as Q$_1$ and Q$_2$, to a polar zone-center mode, Q$_\text{P}$.
The free energy is represented as F $\sim$ Q$_1$ $*$ Q$_2$ $*$ Q$_\text{P}$, illustrating the mutual dependence among these modes \cite{benedek}.
Acting together, both Q$_1$ and Q$_2$, as zone-boundary modes, facilitate the phase transition leading to a polar ground state. 
Thus, both Q$_1$ and Q$_2$ function as the principal order parameters in this scenario.
For both A-site layered DPOs (in $P2_1$ symmetry group) and SLs, the macroscopic origin of the polarization is the non-cancellation of layered polarization in two successive AO and A$^{\prime}$O layers, as established by the HIF mechanism.
The trilinear coupling for these systems is between the in-phase rotation (Q$_{\text{R+}}$), tilt (Q$_{\text{T}}$) and antiferroelectric displacement (Q$_{\text{AFEA}}$).
\par
Figure ~\ref{fig:modes} (a) illustrates the high symmetry, low symmetry phases along with the key structural modes. 
The Q$_{\text{R+}}$ mode vectors represent the clockwise/counter-clockwise rotations of the in-plane O atoms located at the top and bottom layers of 
BO$_6$ (or B$^{\prime}$O$_6$) octahedra.
The movement of apical O atoms located at both layers is considered within Q$_{\text{T}}$ whereas the antiferroelectric displacement of the A-sites, primarily responsible for the layered polarization is represented by Q$_{\text{AFEA}}$.
Recent studies \cite{GPswCM1, GPswCM2} have revealed a two-step thermal switching process facilitated by the out-of-phase rotation Q$_{\text{R-}}$, which exhibits the lowest E$_\text{s}$ compared to the pathways involving the two primary order parameters in HIF DPOs and SLs.
The mechanism points out how $\Vec{P}$ switches from [010] to [0$\bar{1}$0] via Q$_{\text{R-}}$ (a$^{0}$a$^{0}$c$^{-}$ when Q$_{\text{T}}$ = 0 or, a$^{-}$a$^{-}$b$^{-}$ when Q$_{\text{T}}$ $\neq 0$).
We have employed the switching pathway (Figure ~\ref{fig:modes} (b)) via Q$_{\text{R-}}$ to estimate E$_\text{s}$ by performing nudged elastic band calculations.
The composition space considered in this study is shown in Figure ~\ref{fig:modes} (c).
\par
We acknowledge that the chemical space we explore is somewhat restricted compared to the vast array of structures that could be computationally designed by combining elements from the periodic table, adhering to criteria for forming DPOs and SLs. 
This exploration extends to over 59,000 potential structures, with approximately 30,000 structures expected to fall within the monoclinic space group\cite{bare2023dataset}
{\color{black}However, many of these systems may lack the stable A-site cation ordering necessary to demonstrate functionalities such as polarization attributed to HIF mechanism.}
Hence, we focus on the systems that are most likely to form A-site layered ordering.
The details of the first-principles computations are provided in the Methods section.
\par 
\subsection{Causal reasoning}
From the first-principles simulations we curate the features space including information on radii of A-site cations (r$_A$, r$_A^\prime$), difference in radii (r$_\text{diff}$), tolerance factor ($\tau$), displacement of A-site cations (A$_\text{dis}$, A$^\prime_\text{dis}$, rotation angle ($\theta_{r}$), tilt angle ($\theta_{t}$), Kohn-Sham energy (E$_\text{KS}$).
Information on structural modes that have been reported\cite{GPswCM2,ghosh2022insights} to play roles in functionalities in HIFs (e.g.,  Q$_{\text{R+}}$, Q$_{\text{T}}$, Q$_{\text{AFEA}}$, Q$_{\text{AFEO}}$, Q$_{\text{JT}}$, Q$_{\text{CD}}$) are also incorporated into the initial feature space.
The Q$_{\text{AFEO}}$, Q$_{\text{JT}}$, Q$_{\text{CD}}$ signifies the movement of planar O atoms, Jahn-Teller distortion and charge disproportionation via the proportionate change in the bond lengths of B$-$O and B$^\prime-$O, respectively.
\par
All compounds in the DPO and SL dataset are used to construct a causal model with all features. 
The model is then pruned by retaining only the causal connections with strengths above the 90th percentile.
In this model, the sign of the coefficient indicates the direction of the causal relationship: a positive sign signifies a direct causal effect (where an increase in one feature causes an increase in another), while a negative sign indicates an inverse causal effect (where an increase in one feature leads to a decrease in another). 
To ensure all features are comparable on an equal footing, we standardize the dataset using Z-scores. 
This pre-processing step normalizes the data, allowing for a meaningful comparison of the features and their respective impacts on the causal model.
Next, we discard a few features based on the domain-specific context.
%
\begin{figure*}
\includegraphics[width=\textwidth]{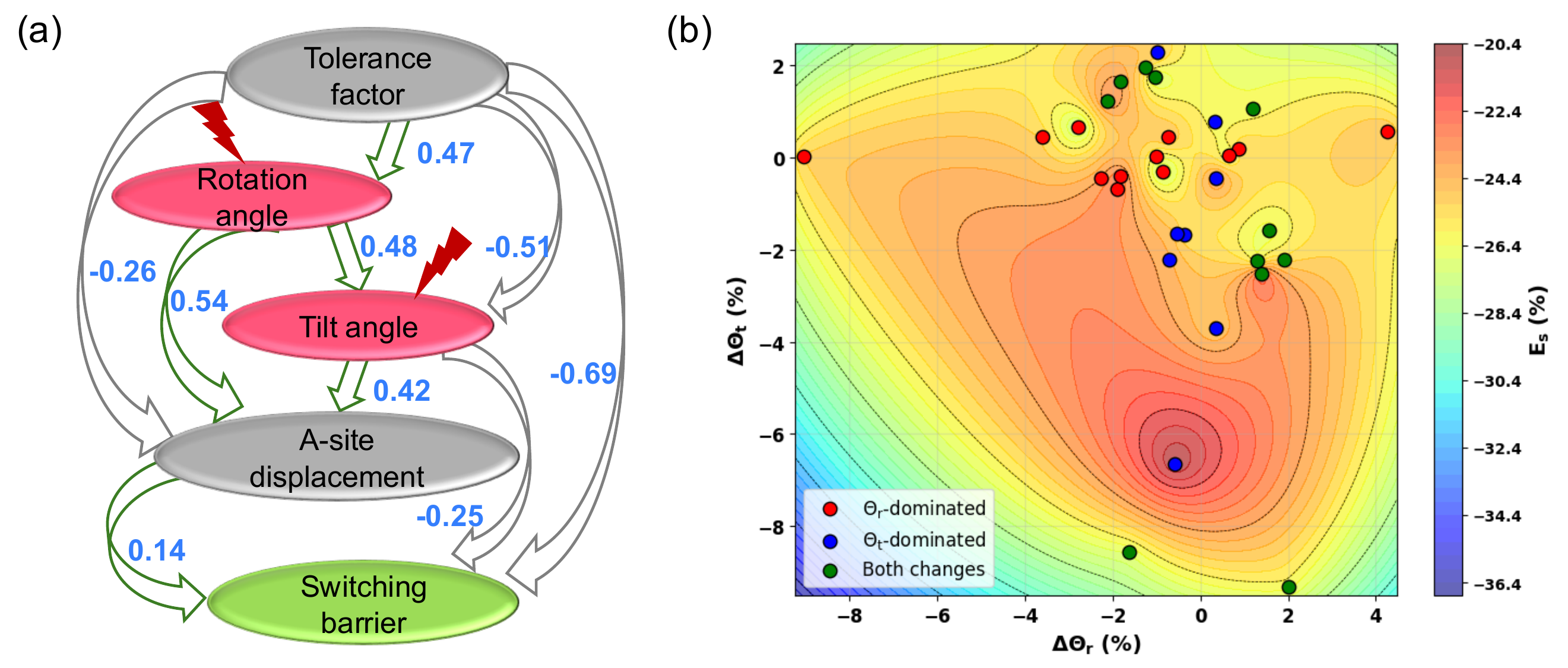}
\caption{(a) Directed acyclic graph illustrating the cause-effect relations between various features and the target. (b) {\color{black}Two-dimensional contour representation of the changes in rotation and tilt that lead to changes in the switching barrier, as estimated by the causal interventions conducted across all materials with $\tau$ $>$0.8 (Group I in our study).}
{\color{black} All materials exhibit negative changes, indicating a reduction in the E$_\text{s}$. Upon intervention, three distinct response pathways are identified:
(1) concurrent and complementary variations in the rotation and tilt angles,
(2) an increase in $\theta_{t}$ with $\theta_{r}$ remaining approximately constant, and
(3) a decrease in $\theta_{r}$ with $\theta_{t}$ remaining approximately constant.
These pathways correspond to distinct distortion mechanisms that reduce $E_{\text{s}}$. The materials following each pathway are denoted in green, blue, and red, respectively.}}
\label{fig:model}
\end{figure*}
%
{\color{black} Although information on structural modes is important, the mode amplitudes are often difficult to characterize and quantify accurately.
Hence, we focus on features that can be more easily modified or controlled in computational models, theoretical analyses, or experimental settings.
We quantify octahedral distortions using $\theta_{r}$ and $\theta_{t}$ angles rather than symmetry-mode amplitudes. 
While the mode amplitudes are rigorously defined in group-theoretical terms, the angles can be readily extracted from crystallographic refinements (X-ray or neutron diffraction), making them experimentally accessible descriptors of octahedral distortions in ABO$_3$ perovskites \cite{fowlie2019,liaopnas2018,qianacsnano2015}. 
The angles, therefore, serve as physically intuitive proxies for the corresponding mode amplitudes that are used as features in our study.
}
\par
The resulting model is shown in Figure~\ref{fig:model} (a).
All the models prior to pruning and selection are provided in the Supplementary Information.
By prioritizing these more accessible features, we ensure that the model remains practical and adaptable, while enabling novel insights.
The refutability tests are performed (results in the Supplementary Information) to ensure the validity of the new model. 
\par
The causal model provides information on how structural parameters influence E$_\text{s}$ in DPOs and SLs. 
The geometric stability of the perovskite structures, as described by $\tau$, has a significant direct negative effect on E$_\text{s}$. 
{\color{black}It suggests that as $\tau$ increases, the perovskite structure becomes more stable, potentially reducing E$_\text{s}$ for polarization switching by allowing for switching by allowing the lattice to accommodate cooperative octahedral distortions (rotation and tilt) more easily under strain. 
This structural flexibility arises when the tolerance factor approaches unity or when A-site size mismatch provides sufficient free volume for distortion.}
The angular distortions ($\theta_{r}$, $\theta_{t}$) of the BO$_6$ and B$^\prime$O$_6$ octahedra, which in turn affect the displacements of the A-site cations, are also strongly modulated by $\tau$.
The causal relations between A$_\text{dis}$ with $\theta_{r}$, $\theta_{t}$ indicate that the octahedral distortions contribute to the off-center shifts of the A-site cations. 
{\color{black}The displacement itself has a small positive effect on E$_\text{s}$, suggesting that larger cation displacements may slightly hinder polarization switching.
The systems with lower $\tau$ exhibit reduced free volume and stronger ionic packing constraints, leading to a more rigid lattice framework. 
Such lattice rigidity suppresses octahedral distortions, resulting in a higher barrier for distortion-driven switching.
}
The $\theta_{t}$ also has a direct negative impact on E$_\text{s}$, implying that an increased octahedral tilt may facilitate switching by modifying the energy landscape.
Although these findings align with current material physics understanding, the precise mechanism by which E$_\text{s}$ is lowered remains, and consequently, it is unclear how polarization switching can be enhanced. 
Lowering E$_\text{s}$ is crucial to improve ferroelectric performance, enabling applications of these materials in devices.
%
\begin{figure}
\centering
\includegraphics[width=\textwidth]{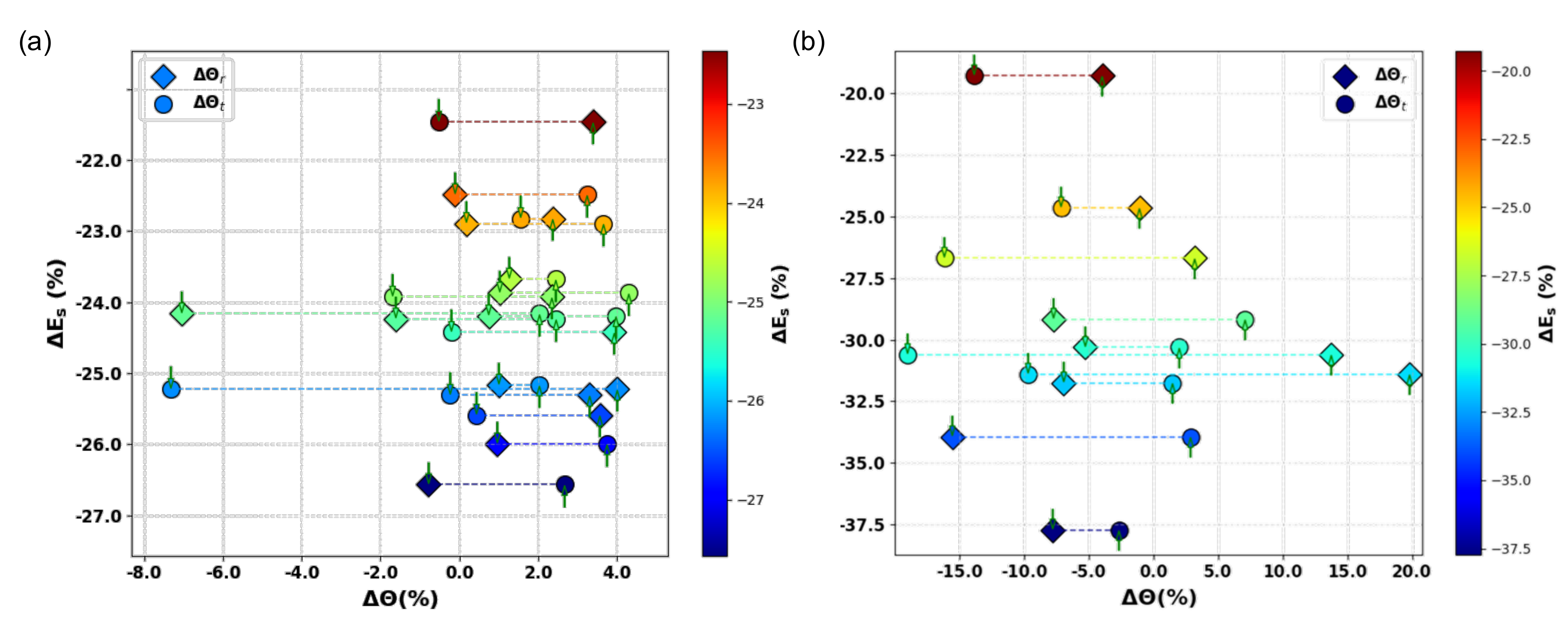}
\caption{Selected Group I materials ($\tau > 0.8$) for which complementary changes in rotation and tilt angles reduce the switching barrier. The required distortions are identified via causal interventions, while DFT calculations provide validation of these material-specific pathways.}
\label{fig:intervention}
\end{figure}
\par
To elucidate the mechanisms that govern polarization switching, we perform causal interventions on $\theta_{r}$ and $\theta_{t}$ in a systematically controlled manner.
Typically when an intervention is performed on a variable ($\theta_{r}$ and $\theta_{t}$ in this case), the edges pointing to that variable (its causes) are removed. 
However, the edges pointing from the intervened variable (its effects) remain intact, allowing us to still observe the causal influence of the intervened variable on other variables in the DAG, effectively isolating it from its usual causes and consequences.
This change allows us to analyze the causal effects of the intervention while controlling for external influences that would otherwise confound the results.
We choose to perform interventions in the \textit{sequential} mode in which changes are applied incrementally to one feature at a time, with each modification influencing subsequent steps, ensuring a gradual and context-sensitive refinement of the model. 
The intervention is constrained within a predefined range, incorporating domain expertise and physics-based principles.
For each selected feature, a random variation of up to $\pm$ 35\% is introduced, ensuring that the changes remain within the observed range of ($\theta_{r}$ and $\theta_{t}$). 
An optimal intervention is then estimated using the pre-built causal model, which predicts the necessary adjustment to achieve a target outcome.
Here for each material, we have placed an upper bound of -50\% change to E$_\text{s}$.
An adaptive weighting approach is applied with a predefined threshold of 0.6: if the required change in $\theta_{r}$ or $\theta_{t}$ is small (below the threshold), the original value is weighted more heavily; for larger adjustments, the intervened value is given a greater weight. 
This approach ensures a gradual modification of the angular distortions while preserving the material structures along with compositions.
The details on the intervention approaches can be found in the Methods section.
\par
\textcolor{black}{
We have selected the compounds with E$_\text{s}$ higher than 30.468 meV/f.u. (median across dataset) for performing interventions to further reduce E$_\text{s}$.}
{\color{black}
Based on the range of $\tau$, we initially divide the filtered dataset into three groups: Group I ($\tau$ $>$ 0.8, mostly DPOs) with 31 compounds, Group II (0.4 $<$ $\tau$ $<$ 0.8, mix of DPOs and SLs) with 36 compounds and Group III ($\tau$ $<$ 0.4, mostly SLs) with 79 compounds.
While our dataset includes systems with $\tau$ $<$0.4, such compositions are unlikely to form stable perovskite structures in practice due to large size mismatch in the cations. 
To the best of our knowledge, no experimentally realized DPOs or SLs exist in this regime. 
Hence, we have focused on the experimentally accessible $\tau$ range of $\sim$0.8-1.0. 
The insights obtained from the causal reasoning are mostly relevant to this range of $\tau$ that includes numerous synthesized ABO$_3$ oxides and its derivatives. 
Results for $\tau$ $<$ 0.4 can be regarded as theoretical extrapolations without immediate experimental relevance.
The intervention results are quantified across all the groups (see Supplementary Information).}
Our discussion primarily focuses on Group I because they are likely to exhibit enhanced structural stability while maintaining the necessary octahedral flexibility for polarization switching.
{\color{black}
\par
Figure~\ref{fig:model}(b) presents the changes in features along with the target values after interventions for Group I.
{\color{black}
After intervention, three types of material-specific distinct changes in the responses are observed, all leading to lowering E$_\text{s}$: (1) both angles change in a complementary fashion (increase or decrease),
(2) $\theta_{t}$ increases at nearly constant $\theta_{r}$ , or
(3) $\theta_{r}$ decreases at nearly constant $\theta_{t}$.
}
Each of these behaviors can be driven by intrinsic lattice dynamics or by external factors such as strain, influencing how the system navigates the energy landscape for polarization reversal.  
In the first case, where the change in angles are opposite, an increased change in one of the angles naturally suppresses the other to maintain structural stability. 
This behavior can often be linked to strain effects, particularly in epitaxial thin films where biaxial strain (compressive or tensile) forces the lattice to adjust by balancing octahedral distortions in a complementary manner.
Strain may also influence the A-site cation displacement, modifying the local electrostatic potential to further reinforce the need for compensatory rotation-tilt adjustments.  
Interestingly, this finding is similar to the behavior captured in the energy landscape of HIFs, where the trilinear coupling is a well-established concept.
By varying specific structural distortions (+/+, +/-, -/+, -/-) that represent the modes involved in trilinear coupling, four equivalent structural minima emerge. 
These drive the system to its ground state as described by the Landau free energy expansion. 
These minima correspond to symmetry-related states that are determined by the underlying crystallographic features, governing the interplay with octahedral rotations or tilts. 
Such factors often influence formation of domain structures and the energetics.
The causal interventions suggests that they also play a crucial role in altering the switching barrier, offering new insights into how they can be used to tailor switching pathways for specific materials as presented in Figure~\ref{fig:intervention} (a). The DFT computed change and in the rotation and tilt angles and reduction in the switching barriers are in agreement with the causal intervention as shown in Figure~\ref{fig:intervention} (b).
\par
When both angles  ($\theta_{r}$ or $\theta_{t}$) increase simultaneously, it may lead to larger off-center cation displacements, effectively strengthening the layered polarization while reducing E$_\text{s}$. 
Strain can also play a role here, particularly if epitaxial constraints stabilize a highly distorted phase, favoring cooperative distortions that promote polarization switching through an energy-lowering mechanism.  
Conversely, a decrease in both angles indicate a possible transition towards a more symmetry-restored phase in which there are reduced or similar distortions as experienced by the system in its lowest energy ground-state phase, thereby fostering an easier switching pathway.
This effect can be strain-driven, as compressive strain can suppress octahedral tilting, bringing the system closer to a centrosymmetric-like structure that allows for smoother rotation of the BO$_6$-B$^\prime$O$_6$ octahedra. 
In addition, lowering of distortion may enhance structural flexibility to make it easier for the material to undergo polarization reversal with a lower energy cost.  
%
\begin{figure*}
\includegraphics[width=\textwidth]{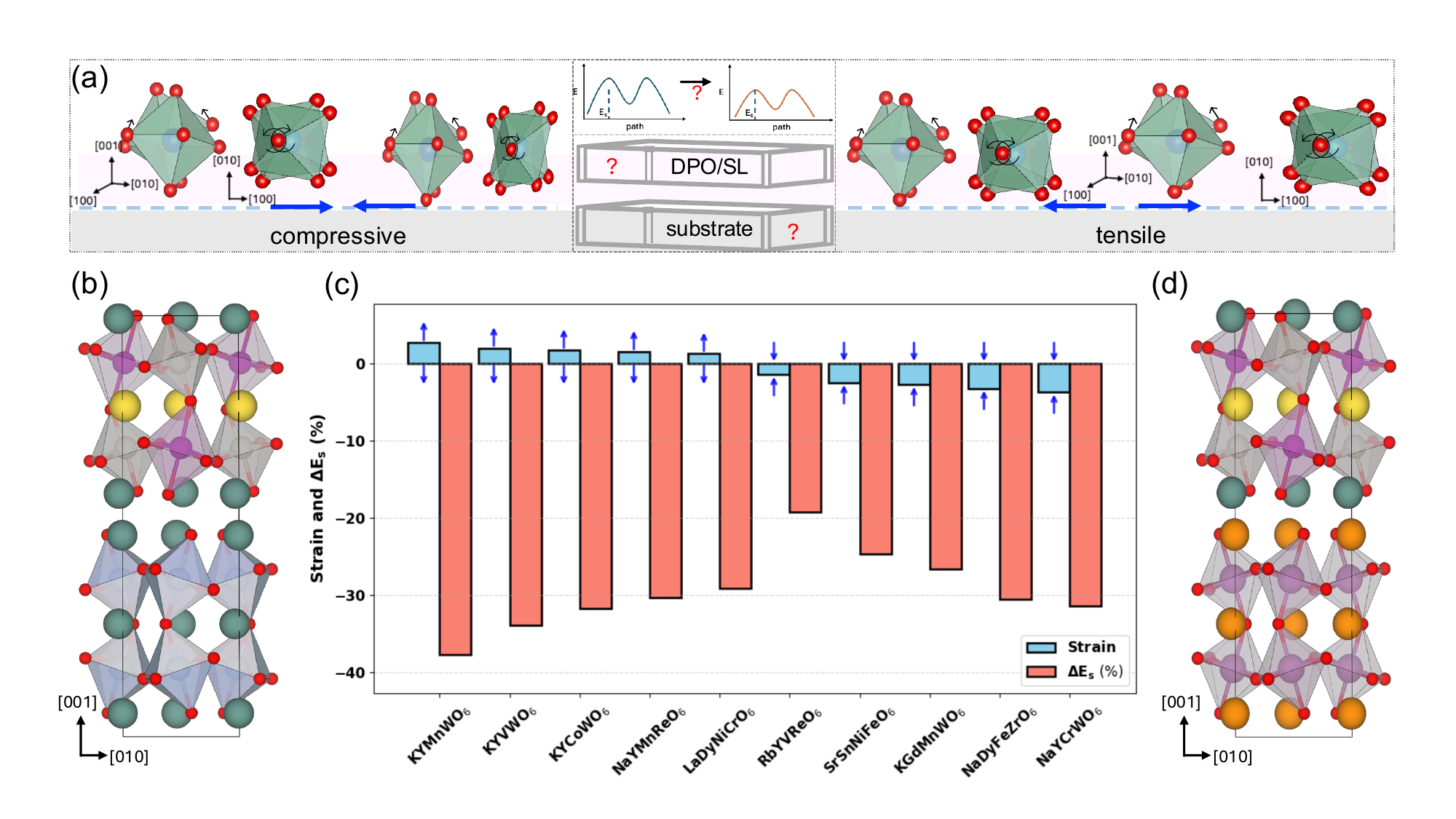}
\caption{{\color{black}(a) Illustration of strain-engineering strategy used to probe switching energetics in perovskites. Substrate-induced biaxial strain produces systematic changes in octahedral rotation and tilt, and each material–substrate pair exhibits a characteristic structural response. Structural models of example substrate/DPO pair for compressive (NaYMnReO$_6$/YAlO$_3$) and tensile (NaYMnReO$_6$/NdScO$_3$) strain are shown in (b) and (d) respectively. (c) Bar plot summarizing the DFT-computed changes in $E_{\text{s}}$ for each material at the tensile or compressive strain that produces the maximal barrier reduction. The causal intervention identifies the required changes in rotation and tilt angles, and epitaxial strain serves as the physical mechanism to enact these distortions. The DFT results validate the intervention-suggested pathways, revealing material-specific mechanisms for modifying the target property.}}
\label{fig:sim_results}
\end{figure*}
%
\par
These distinct structural responses are delineated by the causal mechanisms offering a quantitative understanding of controlling E$_\text{s}$ for a variety of oxide perovskites.
It is intriguing to observe how the mechanisms differ even within the same group, despite having a similar range of $\tau$, suggesting that the behavior is material-specific. 
Such complexity underscores why a generalized mechanism is insufficient, as it cannot capture the full range of material-dependent responses. 
This is precisely why causal intervention is essential—by isolating and manipulating specific variables, we gain a clearer, more nuanced understanding of the underlying mechanisms for each material.
Our hypothesis that the suggested causal mechanisms are strain-induced stems from the fact that these materials are often grown on substrates, which can introduce strain, altering the structural distortions.
The strain effects can also be more readily realized in practice, facilitated by the layer-by-layer growth of A-site layered ordered DPOs and the layered structural arrangements of SLs.}
\par
\textcolor{black}{
To further capture this counterintuitive cooperative pathway, we propose a model of the switching barrier as a function of rotation, tilt, and epitaxial strain.
\[
E_{\mathrm{s}} = E_{\mathrm{s}}(\theta_{r},\theta_{t};\varepsilon).
\]
Expanding around a distorted reference state
$(\theta_{r}^{0},\theta_{t}^{0})$ at fixed strain $\varepsilon$ and
defining $\Delta\theta_{r} = \theta_{r}-\theta_{r}^{0}$ and
$\Delta\theta_{t} = \theta_{t}-\theta_{t}^{0}$, we obtain
\begin{equation}
E_{\mathrm{s}} \approx E_{\mathrm{s}}^{0}
+ \frac{1}{2} k_{r}(\varepsilon)\,(\Delta\theta_{r})^{2}
+ \frac{1}{2} k_{t}(\varepsilon)\,(\Delta\theta_{t})^{2}
+ k_{rt}(\varepsilon)\,\Delta\theta_{r}\Delta\theta_{t}
+ \dots ,
\label{eq:Es_expand}
\end{equation}
where $k_{r}>0$ and $k_{t}>0$ are the self–stiffness of rotation and tilt,
and $k_{rt}$ encodes their cooperative coupling (all re-normalized by strain).}
\textcolor{black}{
The cooperative pathway in which both rotation and tilt increase
simultaneously can be parameterized as $\Delta\theta_{r} = \delta$, $\Delta\theta_{t} = s$,$\delta$, $\delta>0$, $s>0$, such that along this path the barrier becomes
\begin{equation}
E_{\mathrm{s}}(\delta;\varepsilon)
\approx E_{\mathrm{s}}^{0}
+ \frac{1}{2}\,\kappa_{\mathrm{eff}}(\varepsilon)\,\delta^{2}
+ \dots ,
\end{equation}
with an effective curvature
\begin{equation}
\kappa_{\mathrm{eff}}(\varepsilon)
= k_{r}(\varepsilon)
+ k_{t}(\varepsilon)s^{2}
+ 2\,k_{rt}(\varepsilon)s
\label{eq:kappa_eff}
\end{equation}
along the cooperative direction $(\Delta\theta_{r},\Delta\theta_{t}) \propto (1,s)$.
The counterintuitive regime identified by the causal interventions and DFT
corresponds to $\kappa_{\mathrm{eff}}(\varepsilon) < 0$
for the strain values imposed by orthorhombic substrates such as
NdScO$_3$ (tensile) and NdGaO$_3$ (compressive). In this case,
increasing both $\theta_{r}$ and $\theta_{t}$ along the cooperative path
($\delta>0$) \emph{lowers} $E_{\mathrm{s}}$, reflecting a strain-enabled
cooperative coupling in which the energy penalty of enhancing one
distortion is compensated by the other, thereby flattening the energy
landscape for polarization reversal.}
\par
\textcolor{black}{
This non-trivial mechanism, not anticipated by Landau theory, illustrates the power of causal reasoning in identifying new, physics-driven switching mechanisms.
As a result, it enables the identification of experimentally accessible design rules, bridging Landau theory with composition- and strain-specific insights while offering a more comprehensive understanding of how polarization switching can be engineered in HIFs.
}
\subsection{Causal mechanisms as drivers of physics simulations}
Building on this understanding, we then aim to validate the connection between these causal mechanisms with first-principles calculations to provide a detailed understanding of how strain, octahedral distortions, and energy barriers are interrelated.
\par
{\color{black}
Figure~\ref{fig:sim_results} (a) illustrates how causal interventions combined with DFT calculations reveal material-specific mechanisms for tuning the switching barrier.}
We rank the lattice
mismatch between Group I compounds with respect to 11 substrates which are commonly
utilized in experimental layer-by-layer growth.
The analyses include both tensile and compressive strains.
Among the widely used orthorhombic substrates, NdScO$_3$ imposes a tensile strain (ranging from 2.7\% to 1.3\%) on KYMnWO$_6$, KYVWO$_6$, KYCoWO$_6$, NaYMnReO$_6$, and LaDyNiCrO$_6$, whereas NdGaO$_3$ imposes a compressive strain (ranging from 3.7\% to 1.4\%) on RbYVReO$_6$, SrSnNiFeO$_6$, KGdMnWO$_6$, NaDyFeZrO$_6$, and NaYCrWO$_6$.
We find that $\theta_{r}$ increases and $\theta_{t}$ decreases under compressive strain.
In contrast, tensile strain leads to the opposite effect.
{\color{black}
Example substrate-DPO/SL pairs with compressive ((NaYMnReO$_6$/YAlO$_3$) and tensile (NaYMnReO$_6$/NdScO$_3$)) strain are shown in Figure~\ref{fig:sim_results} (b). 
The effects of strain on E$_\text{s}$, as well as changes in structural features, are quantified and compared in Figure~\ref{fig:intervention}(c) with the trends predicted by causal interventions.}
The analysis identifies the required adjustments in octahedral rotation and tilt, while epitaxial strain acts as the experimental handle to realize these distortions. 
The results further validate the intervention-predicted pathways, illustrating how strain can systematically modulate polarization-switching energetics across different perovskite materials.
{\color{black}
\par
Out of the three identified mechanisms, those characterized by either a reduction in rotation at nearly constant tilt or an increase in tilt at nearly constant rotation align with Landau theory predictions, wherein lowering the amplitude of a non-polar distortion reduces the energy barrier. 
The other pathway, however, is counterintuitive: both rotation ($\theta_{r}$) and tilt ($\theta_{t}$) increase simultaneously while the switching barrier decreases.
The causal intervention analysis from our study shows that this outcome arises from a cooperative coupling between the two non-polar modes, in which the penalty in energy to increase a distortion is compensated by the other. 
Under epitaxial strain, as imposed by orthorhombic substrates such as NdScO$_3$ (tensile) and NdGaO$_3$ (compressive), the lattice adopts distortion patterns where increased tilt accommodates increased rotation, thereby flattening the energy landscape for polarization reversal. 
DFT simulations confirm this mechanism for specific DPOs (e.g., KYMnWO$_6$, NaYMnReO$_6$, and RbYVReO$_6$), demonstrating that cooperative distortion tuning can reduce the barrier even when both tilt and rotation amplitudes are enhanced.}
\section{Conclusions}
{\color{black}
Understanding polarization switching in hybrid improper ferroelectrics is a challenging problem due to the complex, interdependent structural and electronic parameters. 
Addressing this challenge requires modern, mechanism-driven approaches, since traditional simulation- and experiment-based methods, though informative, can be limited in identifying generalizable, actionable mechanisms. 
In this work, we introduce a causal reasoning-based approach combined with physics simulations to identify and validate the factors controlling the switching barrier across a chemically diverse set of A-site layered double perovskites and superlattices.
Our analysis reveals that moderate-low intrinsic barriers occur for compositions with tolerance factors between 0.8-1.0 with moderate A-site ionic-radius mismatch, that stabilize layered A/A$^\prime$ ordering while providing sufficient free volume for cooperative octahedral distortions.
The causal intervention identifies three distinct rotation–tilt mechanisms such as complementary changes in rotation and tilt, tilt-dominated changes at nearly fixed rotation, and rotation-dominated changes at nearly fixed tilt, each of which reduces the barrier with the active pathway depending on the specific A/B-site chemistry.
Double perovskite oxides such as KYMnWO$_6$, KYVWO$_6$, 
KYCoWO$_6$, NaYMnReO$_6$, LaDyNiCrO$_6$, RbYVReO$_6$, SrSnNiFeO$_6$, KGdMnWO$_6$, NaDyFeZrO$_6$, and NaYCrWO$_6$ illustrate how appropriate combinations of alkaline-earth/rare-earth A/A$^{\prime}$ ions and 
$3d$--$4d/3d$--$5d$ A/A$^{\prime}$ pairs can sustain large yet switchable octahedral distortions while remaining insulating.
Importantly, this analysis reveals a counterintuitive cooperative mechanism in which simultaneous changes in both rotation and tilt lower the barrier, a pathway that would have been difficult to anticipate from conventional Landau-type arguments alone.
Guided by these causal insights, we demonstrate that epitaxial strain from orthorhombic substrates provides an experimentally accessible knob to realize each material-specific mechanism. 
Tensile strain from NdScO$_3$ and compressive strain from NdGaO$_3$ selectively tune rotation and tilt in ways that reduce the switching barrier by tens of percent, in quantitative agreement with the causal intervention predictions.
For each composition, the analysis prescribes which distortion to enhance or suppress combined with the type of strain to apply, establishing a direct link between substrate choice, octahedral distortion pattern, and barrier.
By combining causal reasoning with first-principles insights, this work not only advances fundamental understanding of complex ferroelectric switching behavior but also provides experimentally actionable guidelines for designing low-barrier hybrid improper ferroelectrics, with potential applications in spintronics and beyond.
More broadly, our approach is generalizable to other complex oxides or functional materials, where uncovering the underlying physics is essential for guiding rational design to accelerate materials discovery.
}
{\color{black}
\section {Supporting Information}
The supporting information includes information on distribution of (a) key features, (b) groups of compounds based on tolerance factor of the full dataset, causal maps including (c) all, (d) selected features, (e) outcome of refutability tests, (f,g) quantification of changes in tilt and rotation distortions for different group of compounds
}
\section {Acknowledgments}
This research (A.G.) was sponsored by the Laboratory Directed Research and Development Program of Oak Ridge National Laboratory, managed by UT-Battelle, LLC, for the U. S. Department of Energy. ORNL is managed by UT-Battelle, LLC, for DOE under Contract No. DE-AC05-00OR22725. Theoretical analyses (A.G.) were funded by U.S. Department of Energy, Office of Science, Office of Basic Energy Sciences, Materials Science and Engineering Division. S.G. acknowledges funding from DST-SERB, Core Research Grant, Ref. CRG/2023/3209, Government of India.
\section{Author Contributions}
A.G. developed the idea, performed calculations and wrote the manuscript. P.G. and S.B. participated in performing computations.
A.G. and S.G. together developed the understanding of the underlying physics.
\section {Conflicts of Interest}
The authors have no conflict of interest to declare.
\section {Data Availability}
The datasets utilized in this study can be found in the publicly available \href{https://github.com/aghosh92/causal-modeling-phys}{Github repository}.
\section {Code Availability}
The causal reasoning worklfow and results as mentioned in the study can be found in \href{https://github.com/aghosh92/causal-modeling-phys}{Github repository}.
\label{PST}
\providecommand{\latin}[1]{#1}
\makeatletter
\providecommand{\doi}
  {\begingroup\let\do\@makeother\dospecials
  \catcode`\{=1 \catcode`\}=2 \doi@aux}
\providecommand{\doi@aux}[1]{\endgroup\texttt{#1}}
\makeatother
\providecommand*\mcitethebibliography{\thebibliography}
\csname @ifundefined\endcsname{endmcitethebibliography}
  {\let\endmcitethebibliography\endthebibliography}{}

\end{document}